\RequirePackage[T1]{fontenc}

\documentclass[alpha-refs, onecolumn, doublespace]{wiley-article}

\usepackage{setspace}

\usepackage{lmodern}
\AtBeginDocument{}

\usepackage{anyfontsize} % allows arbitrary font sizes

\usepackage{threeparttable}
\usepackage{soul,color,ulem} % for highlighting text
\usepackage{lineno}

\usepackage{graphicx}
\usepackage{longtable}
\usepackage{tabulary}
\usepackage{booktabs,array,multirow}
\usepackage{amsfonts,amsmath,amssymb}
\usepackage{natbib}
\usepackage{hyperref}
\hypersetup{colorlinks=false,pdfborder={0 0 0}}
\usepackage{etoolbox}


\newif\iflatexml\latexmlfalse

\AtBeginDocument{\DeclareGraphicsExtensions{.pdf,.PDF,.eps,.EPS,.png,.PNG,.tif,.TIF,.jpg,.JPG,.jpeg,.JPEG}}

\usepackage[english]{babel}

\usepackage{siunitx}

\iflatexml
\else
\corraddress{Umair Bin Waheed, Department of Geosciences, King Fahd University of Petroleum and Minerals, Dhahran 31261, Saudi Arabia}
\corremail{umair.waheed@kfupm.edu.sa}
\fundinginfo{The authors received no specific funding for this work.}
\fi

\papertype{Original Article}

\title{Seismic event classification with a lightweight Fourier Neural Operator model}

\author[1]{Ayrat Abdullin}
\author[1,*]{Umair bin Waheed}
\author[2]{Leo Eisner}
\author[1]{Abdullatif Al-Shuhail}

\affil[1]{King Fahd University of Petroleum and Minerals, Department of Geosciences, Dhahran, 31261, Saudi Arabia}
\affil[2]{Seismik s.r.o., Prague, 18200, Czech Republic}

\runningauthor{Abdullin et al.}

\begin{document}

\maketitle

\begin{abstract}

\textbf{Abstract}

\noindent

Real-time monitoring of induced seismicity is critical to mitigate operational risks, relying on the rapid and accurate classification of triggered data from continuous data streams. Deep learning models are effective for this purpose but require substantial computational resources, making real-time processing difficult. To address this limitation, a lightweight model based on the Fourier Neural Operator (FNO) is proposed for the classification of microseismic events, leveraging its inherent resolution-invariance and computational efficiency for waveform processing.
In the STanford EArthquake Dataset (STEAD), a global and large-scale database of seismic waveforms, the FNO-based model demonstrates high effectiveness for trigger classification, with an F1 score of 95\% even in the scenario of data sparsity in training. The new FNO model greatly decreases the computer power needed relative to current deep learning models without sacrificing the classification success rate measured by the F1 score. A test on a real microseismic dataset shows a classification success rate with an F1 score of 98\%, outperforming many traditional deep-learning techniques. The reduced computational cost makes the proposed FNO model well suited for deployment in resource-constrained, near–real-time seismic monitoring workflows, including traffic-light implementations.
The source code for the proposed FNO classifier will be available at: https://github.com/ayratabd/FNOclass.

\vspace{10pt} % adds 10 points of vertical space

\noindent
\textbf{KEYWORDS}

\noindent
Microseismic, Neural Operators, Machine Learning, Induced Seismicity

\end{abstract}%

% \twocolumn % This command switches the layout to two columns
\doublespacing
% \linenumbers

\section*{INTRODUCTION}

It is well-documented that the injection or extraction of fluids in subsurface geological formations can trigger induced seismicity~\citep{healy1968denver}. Fluid injection typically induces geomechanical instability by reducing the effective normal stress due to increased pore pressure~\citep{zoback2010reservoir}. In addition, induced seismicity may also result from other mechanisms, such as fault reactivation triggered by cooling~\citep{kivi2022cooling}, reservoir compaction \citep{segall1989earthquakes,vlek2019groningen}, or stress redistribution~\citep{kettlety2020stress}. The diverse mechanisms capable of initiating seismic events, combined with uncertainties regarding subsurface conditions, including stress fields, thermal gradients, and fault geometry, result in limited \textit{a priori} predictability of induced seismicity regarding its spatial occurrence, timing, and magnitude.

Such hazards are reduced by seismic monitoring with real-time detection and characterization of induced seismicity during anthropogenic operations~\citep{verdon2021green}. The seismic event characterization requires triggering, classification of triggers, and phase picking, which is a challenge in both earthquake seismology and microseismic monitoring. Traditionally, seismic analysts have determined earthquake signals by visually inspecting three components of waveforms of multiple stations and manually selecting P- and S-wave phase arrivals. However, this method is time-consuming and prone to errors. In particular, these tasks are difficult for microseismic monitoring due to the typically low magnitudes of microseismic events and their susceptibility to noise interference. Visual detection is impractical in real-time operations, and the growing volume of seismic data makes manual analysis impossible, requiring automated algorithms. The most commonly used algorithms for event detection (i.e., triggering and classification) are STA/LTA~\citep{allen1978automatic} and template matching~\citep{gibbons2006detection}. Although STA/LTA provides computational efficiency, its sensitivity to low-magnitude events and time-varying noise is limited, and its performance is somewhat subjective as it depends on the choice of STA/LTA parameters. The subjectivity of STA/LTA detection is overcome by selecting low thresholds in the detection, resulting in a high rate of triggers with false detections. Hence, the triggers need to be classified into true and false detections. Template matching, on the other hand, is robust in detecting small events, but struggles with waveform variability and high computational requirements, making it difficult to apply in real-time~\citep{yoon2015earthquake}. Furthermore, the template matching requires previously recorded events as the templates, which are rarely available at induced seismicity monitoring. 

With the growing interest in machine learning (ML) in the seismological community, significant efforts have been devoted to better utilize it in trigger classification. ML has been used in the classification and detection of seismic triggers since the 1990s and many approaches have been suggested. Many of them use artificial neural networks applied to features extracted from recorded seismic signals \citep{gentili2006automatic,maity2014novel,qu2020automatic}. We note that these algorithms are applied to stationary networks with previously recorded seismicity, which is used as training data. This is usually not possible for induced seismicity where the networks are newly installed to monitor anthropogenic activity. In recent years, several DL models for detection and phase picking have been published \citep{ross2018generalized,mousavi2019cred,woollam2019convolutional,zhu2019phasenet,soto2021deepphasepick}. Existing DL models applied to seismic data analysis differ in several key dimensions, including network architecture, such as convolutional, recurrent, and attention-based frameworks; data input representation, either in the time or frequency domain; and output labeling methods, ranging from discrete point to continuous predictions. These models differ in complexity levels, ranging from relatively simple to deeper architectures.

Various supervised DL methods have been developed, which can be classified based on the type of information they provide, from determining whether an event occurs in a triggered interval of data (classification) to determining the arrival time of the event (picking). Several binary window classification approaches have been developed, in which DL models analyze a time window to predict the presence or absence of an event. For example, \cite{wilkins2020identifying} developed a custom convolutional neural network (CNN) to detect microseismic events associated with mining operations, achieving a ten-fold increase in event detection compared to human experts. Although earlier methods often focused on individual seismic stations, \cite{shaheen2021groningennet} presented a CNN-based approach to detect small-magnitude events using data from an entire network of stations (shallow borehole seismic stations in Groningen, the Netherlands). In addition, \cite{ross2018generalized} proposed a generalized phase detection (GPD) method by training a CNN on manually labeled data from the Southern California seismic network to classify data intervals as containing P-waves, S-waves, or noise. For arrival-time estimation in microseismic settings, transfer-learning and U-Net–type architectures have been used to improve phase picking when labeled microseismic picks are limited~\citep{zhang2022phase}.

Both non-neural and neural network (NN) approaches have shown significant advantages over traditional trace-based detection methods. 
\cite{chen2020automatic} proposed an unsupervised fuzzy clustering method that employs various features such as STA/LTA triggers, interval powers and means to effectively identify microseismic events even when the waveforms are highly noisy. Meanwhile, \cite{zheng2018automatic} and \cite{birnie2022bidirectional} used recurrent long short-term memory (LSTM) network architecture. \cite{zheng2018automatic} trained their LSTM model on laboratory-simulated microseismic events and successfully applied it to field data. \cite{birnie2022bidirectional} further improved the LSTM model by incorporating bidirectionality and training it on various synthetic datasets with band-pass noise. \cite{mousavi2019cred} developed a CNN-RNN earthquake detector (CRED) that combines convolutional layers with bidirectional LSTM blocks in the residual structure to improve microearthquake detection. Using a different approach, \cite{birnie2021introduction} treated detection as a semantic segmentation problem for the array records, where seismic data are represented as images, and the network predicts the presence of an event at each pixel. More recently, \cite{awais2023first} demonstrated that combining super-virtual refraction interferometry with the unsupervised DBSCAN clustering algorithm enables accurate and automatic first-break traveltime picking even in noisy seismic data sets. Beyond detection and picking, DL methods have increasingly been used for rapid microseismic monitoring tasks such as source localization (often trained on synthetic waveforms and applied to field data) and processing of large distributed acoustic sensing (DAS) datasets \citep{wamriew2021deep,vinard2022localizing,yu2024daseventnet}.

Neural operators, as extensions of neural networks, are very effective in constructing surrogate models for various scenarios governed by partial differential equations (PDEs)~\citep{kovachki2023neural}. The Fourier neural operator (FNO)~\citep{li2020fourier}, in particular, excels at training and generalizing PDEs from data. Unlike conventional numerical solvers that solve a single instance of the PDE, the proposed FNO-based approach provides a resolution-invariant solution to the event detection and classification problem, improving both accuracy and efficiency.

Most deep learning models require a high number of trainable parameters on the order of hundreds of thousands. This, in turn, leads to increasing computational resources and time needed to train such a model. Models with millions of parameters are impractical, especially for real-time applications such as microseismic event detection and phase picking. One of our goals in this study is to explore lightweight neural operator models with less computational burden. In the end, reducing the complexity of the computational models leads to less environmental impact, including carbon emissions and pollution.

In this study, we propose a lightweight seismic signal detection model based on FNO. We first apply the FNO framework to classify seismic events from the STanford EArthquake Dataset (STEAD)~\citep{mousavi2019stanford}. We train and evaluate FNO on real earthquake datasets, demonstrating high accuracy. Then we train and test our algorithm on real microseismic data with further improved performance. We also benchmark this method with leading DL algorithms widely used in the seismological community~\citep{zhu2019phasenet,mousavi2020earthquake}. Showing similar performance with PhaseNet, our model has much fewer trainable parameters, which on a large scale, leads to less computation burden both in training and testing large amounts of seismic data. The results show that the use of FNO for detecting seismic events can be useful for near-real time  and cost-effective microseismic monitoring.

\section*{DATA AND METHODS}

\subsection*{Datasets}

For this study, we first used data from the STanford EArthquake Dataset (STEAD)~\citep{mousavi2019stanford} to analyze seismic events. 
STEAD is a large dataset of seismic signals worldwide and encompasses local earthquakes and non-earthquake seismic events (seismic noise) labeled by humans. It includes approximately 1.2 million three-component seismograms, totaling more than 19,000 hours of seismic recordings. The data were acquired by a network of 2,613 seismic instruments around the world, located within a 350 km radius of recorded earthquakes. Each sample consists of 6,000 data points recorded over one minute at a sampling rate of 100 Hz, and three channels: east-west (E), north-south (N), and vertical (Z). 
The dataset is labeled, which is important for developing and testing machine learning algorithms for identifying and classifying earthquakes.

This study also utilizes a microseismic dataset comprising waveforms recorded by a surface array during hydraulic fracturing operations. The acquisition geometry consisted of nine three-component geophones distributed over an area of approximately 100 km², operating at a sampling frequency of 250 Hz. We downsampled the continuous recordings to 100 Hz, as the dominant frequency of the observed seismic signal is below 50 Hz and 100 Hz is the original sampling rate for all models tested. The dataset contains low-magnitude induced seismicity ($0.5 < M < 2.5$) with distinct P- and S-wave arrivals, as well as background noise, which includes non-seismic events specific to the field. From the continuous data, we assembled a balanced set of approximately 1,400 60-sec waveforms by selecting approximately equal numbers of event and noise windows (to prevent class imbalance during training). This compilation includes up to 600 manually labeled local and near-network events (epicentral distances > 10 km from the center of network), supplemented by representative noise intervals. All waveforms were preprocessed with amplitude normalization to standardize inputs for the deep-learning models.

We intentionally refrained from applying synthetic data augmentation techniques, such as superimposing multiple events or inserting zero-value gaps, to preserve the authentic statistical properties of the field recordings. While the dataset naturally contains rare instances of overlapping events and near-zero intervals, we prioritized learning from these intrinsic field characteristics rather than artificially synthesizing complex scenarios.

\subsection*{Pre-trained Models}

We evaluate five models for the microseismic event classification task using their original model weights (Table~\ref{tab:models}). All models are originally trained and tested here on single-station three-component data. The model architectures and weights are available in the SeisBench~\citep{woollam2022seisbench}.

\begin{table*}[htbp]
  \centering
  \resizebox{\textwidth}{!}{%
  \begin{threeparttable}
    \caption{Comparison of the deep learning models}
    \label{tab:models}
    \begin{tabular}{lcccccc}
      \toprule
      & FNO      & CRED      & DPP      & EQT      & GPD      & PhaseNet \\
      \midrule
      \# Params
      & 34{,}000      & 293{,}569      & 199{,}731      & 376{,}935      & 1{,}741{,}003      & 268{,}443 \\
      Type
      & NO      & CNN-RNN      & CNN      & CNN-RNN-Attention      & CNN      & U-Net \\
      Training set
      & Microseismic      & N. California      & INSTANCE      & STEAD      & S. California      & N. California \\
      Station spacing (km)
      & 5-10      & 10-30      & 25-30      & 10-30      & 10-20      & 10-30 \\
      Earthquake depth (km)
      & 1.7-2.4      & 0.0-15.0      & 5.0-15.0      & 0.0-15.0      & 2.0-15.0      & 0.0-15.0 \\
      Earthquake magnitude
      & 0.5-2.5      & 0.5-3.5      & 2.0-4.0      & 0.5-3.0      & 0.0-2.5      & 0.5-3.5 \\
      Orig.\ train size (\# events)
      & 200      & 440{,}000      & 32{,}400      & 1{,}100{,}000      & 3{,}375{,}000      & 623{,}000 \\
      Orig.\ weights
      & Yes      & Yes      & No      & Yes      & Yes      & Yes \\
      Reference
      & This paper      & (Mousavi, Zhu, et al., 2019)      & (Soto \& Schurr, 2021)      & (Mousavi et al., 2020)      & (Ross et al., 2018)      & (Zhu \& Beroza, 2019) \\
      \bottomrule
    \end{tabular}
    \begin{tablenotes}
      \footnotesize
      \item \textbf{Note:} "\# of parameters" refers to the total number of trainable parameters; minor deviations from the original data are possible due to differences in implementation methods. "Training size" indicates the approximate number of examples used to train the original model. "Original weights" indicates whether the weights of the original pre-trained model are available in SeisBench.
    \end{tablenotes}
  \end{threeparttable}
  }
\end{table*}

The CNN-RNN Earthquake Detector (CRED) is designed exclusively for earthquake detection (i.e., trigger and classification) and does not support seismic phase identification or phase picking~\citep{mousavi2019cred}. Inputs into the network are spectrograms of three-component seismograms. CRED analyzes spectrogram representations of seismic waveforms with durations of 30 seconds, sampled at 100 Hz. Its internal architecture integrates convolutional neural network (CNN) layers followed by long short-term memory (LSTM) units, ultimately producing a prediction curve indicating earthquake detections at specific time points within the waveform. The initial training of CRED involved a dataset comprising 550,000 event waveforms and an equal number of waveforms without signal (only seismic noise) collected from Northern California. In this study, CRED was evaluated solely for the earthquake detection task, utilizing the output of the pick detection function.

DeepPhasePick (DPP) is a suite of models designed for earthquake detection and seismic phase picking~\citep{soto2021deepphasepick}. Its detection module employs a convolutional neural network (CNN) structure utilizing depth-wise separable convolutions, which assign probabilities of noise (N), P-phase (P), or S-phase (S) to 5-second waveform time intervals. The intervals to be classified are extracted close to the earthquake origin times and analyst-picked P- and S-phase arrivals, respectively. The model receives these three-component seismic waveform windows as inputs and outputs three-probability vectors corresponding to the predicted class labels (P, S, or N). Originally, DPP was trained on a dataset comprising more than 65,000 waveforms collected around the 1995 Mw = 8.1 Antofagasta and the 2007 Mw = 7.7 Tocopilla earthquakes in Northern Chile. Here, we use the DPP trained on the data from the Italian Seismic Network (INSTANCE)~\citep{michelini2021instance}, as models trained on it perform best on average in all tasks~\citep{munchmeyer2022picker}. To adapt this architecture to our 60 s recordings, we partitioned each waveform into 12 non-overlapping 5 s intervals and aggregated the predictions, such that the entire window was classified as an event if the probability in any sub-interval exceeded the threshold.

Earthquake Transformer (EQTransformer) is a neural network designed for joint earthquake detection, seismic-phase identification, and phase (onsets) picking~\citep{mousavi2020earthquake}. The model operates on waveform segments of 60 s duration sampled at 100 Hz, outputting three probability prediction traces corresponding respectively to earthquake detection, P-phase, and S-phase occurrences at each time point within the window. Internally, EQTransformer integrates convolutional neural networks (CNNs), long short-term memory (LSTM) units, and self-attention layers. There are also various data augmentation techniques used for the training data: adding Gaussian noise, random data gaps, and random channel drops. It also shifts waveform segments in cycles, allowing the phase picks to occur at any point in the window. The model was originally trained on the STEAD dataset~\citep{mousavi2019stanford}. For the binary classification task, we utilized the model's dedicated detection output channel, classifying a window as a positive event if the detection probability exceeded the evaluation threshold at any sample point within the 60 s trace.

Generalized Phase Detection (GPD) is a seismic-phase identification model that analyzes short waveform windows of 4 s duration, sampled at 100 Hz, and classifies each window as P-phase, S-phase, or noise for a single station~\citep{ross2018generalized}. GPD primarily performs phase classification, but applying the trained model in a sliding-window approach enables its use for phase onset detection (similarly to DPP). GPD was originally trained and validated using 4.5 million waveform samples from Southern California, evenly distributed among P-phase arrivals, S-phase arrivals, and noise segments.

PhaseNet is a U-Net-based deep neural network specifically designed for seismic phase arrival-time picking~\citep{zhu2019phasenet}. It processes waveform segments of 30 seconds sampled at 100 Hz, generating continuous probability traces for P- and S-wave arrivals, each of the same duration as the input window. The threshold of probability is set to 0.5 for both P and S picks. Structurally, PhaseNet employs large convolutional filter sizes, incorporates convolutional layers without stride, and utilizes residual connections. The original PhaseNet model was trained and validated on a dataset comprising 779,514 seismic waveforms with labeled P- and S-wave arrivals from Northern California. Similarly to the EQTransformer, we utilized the model's noise output channel, classifying a window as a positive event if the noise probability fell below the evaluation threshold at any point within the trace.

\subsection*{FNO-based Model Training}

We propose a neural operator approach for classifying seismic events. This algorithm uses Fourier Neural Operator layers to process temporal data using fast Fourier transforms, allowing solutions to be computed quickly. The FNO blocks encode the seismogram data efficiently due to their regular sampling nature. A key advantage of neural operators over traditional neural networks is their sample invariance, which allows the input features to be discretized on different grids during each estimation. This eliminates the need for model retraining and allows the algorithm to function efficiently in a variety of seismic recording settings. This property is critical for event detection in dynamic seismic networks, especially for microseismic monitoring, where the acquisition setup may be different for each case study.

\begin{figure*}[!htbp]
\begin{center}
\includegraphics[width=0.9\textwidth]{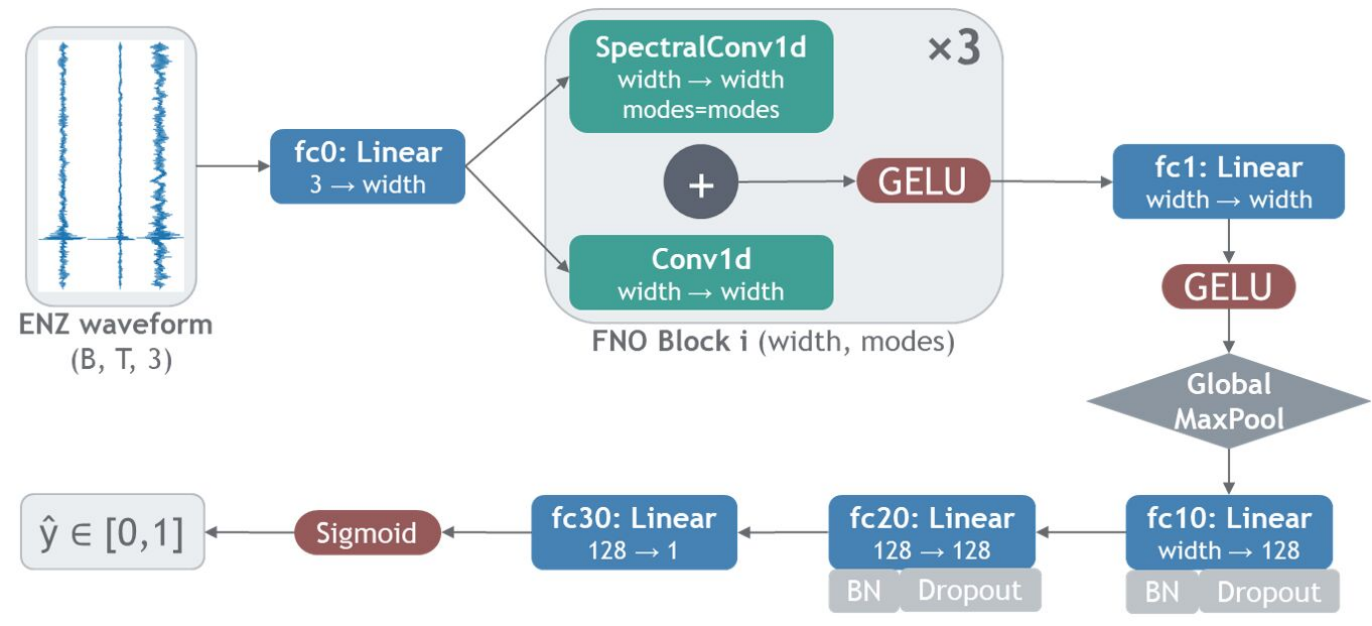}
\caption{
A simple diagram illustrates the proposed Fourier Neural Operator (FNO) method for detecting seismic events. The input is a three-component seismic waveform window (E, N, Z). The input is first transformed into a higher-dimensional feature space, then processed by three connected 1-D FNO blocks. Each block uses a spectral convolution and a local 1x1 convolution, then a nonlinear activation function, to capture both global and local waveform characteristics. The features are then combined by a global max pooling, and passed through a fully connected classifier that outputs a single number, which is the probability that the window contains a seismic signal (1 = signal, 0 = noise).
}
\label{fig:method}
\end{center}
\end{figure*}

We modify the model architecture proposed by \cite{kashefi2024novel} for the seismic event classification. The FNO architecture used in this study consists of three 1-D spectral convolution blocks (Figure~\ref{fig:method}). The FNO blocks capture both global and local waveform characteristics, allowing for efficient detection of seismic events in complex time series data exhibiting varying signal-to-noise ratios. The input is a three-component seismic waveform, and the output is a single scalar, which is the probability that the window contains a seismic signal.

The spectral convolution blocks work in the Fourier domain with two important parameters: \textit{modes} and \textit{width}. The \textit{modes} parameter determines how many Fourier modes are retained in each block, with higher-order modes capturing finer details in the data and lower-order modes focusing more on general patterns. The \textit{width} of each block determines the number of features or channels that the network can capture at each layer, which affects the model’s ability to learn complex relationships in seismic data, such as amplitude and SNR variations, waveform dispersion.

The training process uses the AdamW optimizer with a learning rate of $1 \times 10^{-3}$ along with the ReduceLROnPlateau scheduler to reduce the learning rate when the validation loss plateaus. The loss function is the binary cross-entropy (BCE), which the model seeks to minimize during training. For earthquake detection, the seismic events are randomly divided into training (100 to 500 one-minute waveforms: balanced signal and noise), validation (50 signal and 50 noise samples), and test (500 signal and 500 noise) sets. The microseismic dataset was randomly divided into three subsets: a training set of 200 one-minute waveforms (100 seismic signals and 100 noise segments), an identical validation set, and a larger test set of 1000 waveforms (500 seismic signals and 500 noise segments). The validation set is used to fine-tune the learning rate, while the test set is used to evaluate the model's performance.

\subsection*{Evaluation Metrics}

\begin{figure}[!htbp]
\begin{center}
\includegraphics[width=0.5\linewidth]{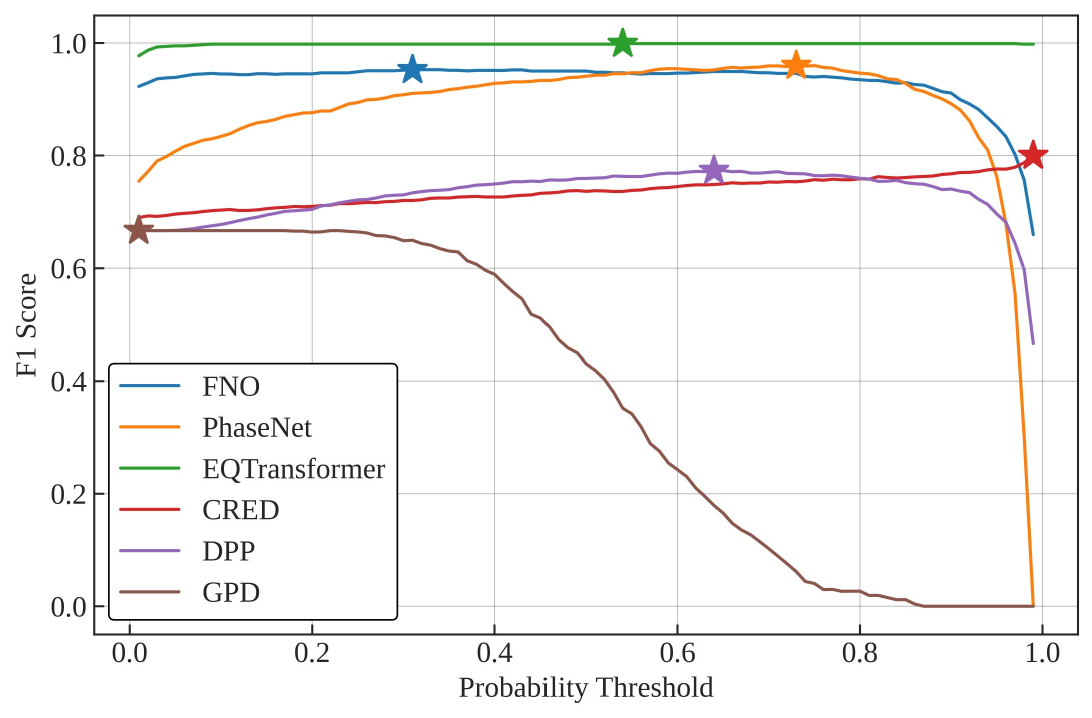}
\caption{
Sensitivity analysis of the F1 score for the STEAD test set with respect to the probability threshold for the six evaluated deep learning models. The curves illustrate the dependency of detection performance on the chosen decision boundary across the full probability range [0, 1]. Star markers indicate the optimal threshold identified to maximize the F1 score for each respective architecture.
}
\label{fig:f1_th_opt}
\end{center}
\end{figure}

Model evaluation is conducted by analyzing the residuals between the ground truth (STEAD or manual labels) and the model's predictions, as well as by computing performance metrics derived from the confusion matrix. Seismic waveforms correctly identified as earthquakes were categorized as true positives (TP), whereas seismic signals incorrectly missed by the model were considered false negatives (FN). Similarly, noise waveforms accurately classified as non-seismic signals were true negatives (TN), and noise signals incorrectly classified as earthquakes were false positives (FP). Based on these results, precision and recall were computed according to Equations~\ref{eq:prec} and~\ref{eq:rec}, respectively. Furthermore, the F1-score, representing the harmonic mean of precision and recall, was determined following Equation~\ref{eq:f1}. The F1 score depends on a choice of decision threshold. For evaluating seismic event detection performance, we chose the threshold probabilities maximizing the F1 score on the test set independently for each method (Figure~\ref{fig:f1_th_opt}).

\begin{align}
    \mathrm{precision} &= \frac{TP}{TP + FP} 
    \label{eq:prec} \\
    \mathrm{recall} &= \frac{TP}{TP + FN} 
    \label{eq:rec} \\
    F_1\text{-score} &= 2 \times \frac{\mathrm{precision} \times \mathrm{recall}}{\mathrm{precision} + \mathrm{recall}} 
    \label{eq:f1}
\end{align}

% \begin{equation}
% \mathrm{precision} = \frac{TP}{TP + FP}
% \tag{1}
% \label{eq:prec}
% \end{equation}

% \begin{equation}
% \mathrm{recall} = \frac{TP}{TP + FN}
% \tag{2}
% \label{eq:rec}
% \end{equation}

% \begin{equation}
% F_1\text{-score} = 2 \times 
% \frac{\mathrm{precision}\,\times\,\mathrm{recall}}
%      {\mathrm{precision} + \mathrm{recall}}
% \tag{3}
% \label{eq:f1}
% \end{equation}

\section*{RESULTS AND DISCUSSION}

\subsection*{STEAD Dataset Performance}

\begin{figure}[!htbp]
\begin{center}
\includegraphics[width=0.5\textwidth]{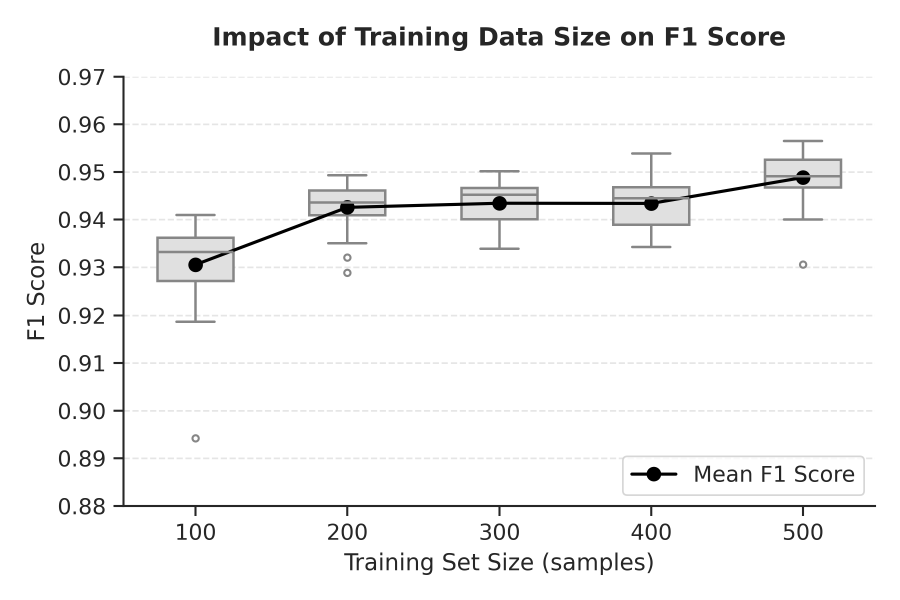}
\caption{
Impact of training dataset size on the classification performance (F1 score) of the FNO model. The box plots illustrate how F1 scores are distributed for different model setups (different modes and widths) for different training set sizes (ranging from 100 to 500 samples). The line in the middle of each box marks the median F1 score. The edges of the box indicate the interquartile range. The line with markers tracks the mean F1 scores, illustrating that the accuracy increases and stability improves (reduced variance) as the number of training samples increases.
}
\label{fig:training_impact}
\end{center}
\end{figure}

To verify the FNO model's effectiveness under data-sparse conditions, we first evaluated its performance across training set sizes ranging from 100 to 500 samples (Figure~\ref{fig:training_impact}). The results indicate that the FNO model maintains a robust F1 score above 0.93 even with minimal training data, with performance stability significantly improving as the dataset size increases beyond 200 samples (Table~\ref{tab:training_size_stats}). For the final performance evaluation and comparison with the results of the other methods, the model trained on the largest dataset (500 samples) was used. The hyperparameters of the model were selected based on the lowest validation loss: modes = 16, width = 32.

\begin{table}
\centering
\caption{Performance statistics (F1 score) of the FNO model across varying training dataset sizes.}
\label{tab:training_size_stats}
% This command forces the table to scale down to the column width
% \resizebox{\columnwidth}{!}{%
    \begin{tabular}{ccccc}
    \toprule
    Training Size & Mean F1 & Std Dev & Min F1 & Max F1 \\
    \midrule
    100 & 0.9305 & 0.0093 & 0.8942 & 0.9409 \\
    200 & 0.9426 & 0.0050 & 0.9288 & 0.9493 \\
    300 & 0.9434 & 0.0047 & 0.9339 & 0.9501 \\
    400 & 0.9434 & 0.0053 & 0.9342 & 0.9539 \\
    500 & 0.9488 & 0.0054 & 0.9306 & 0.9565 \\
    \bottomrule
    \end{tabular}%
% }
\end{table}

For the testing of the model’s ability to generalize, a blind test was conducted on 1,000 independent waveform windows ($N_{noise}=500$, $N_{signal}=500$). Table~\ref{tab:model_performance} shows the performance metrics for the FNO model and the five different architectures. The FNO-based model’s accuracy was 95.2\%, and the F1 score was 0.953. The highest metrics were for the EQTransformer and the PhaseNet models, with F1 scores of 0.999 and 0.960, respectively. It is important to note that these baseline models benefit from extensive pre-training on large-scale global datasets. However, the FNO model shows a strong ability to detect the main features of the waveforms and deliver high-quality results with a limited training set of only 500 samples.

\begin{figure}[!htbp]
\begin{center}
\includegraphics[width=0.5\linewidth]{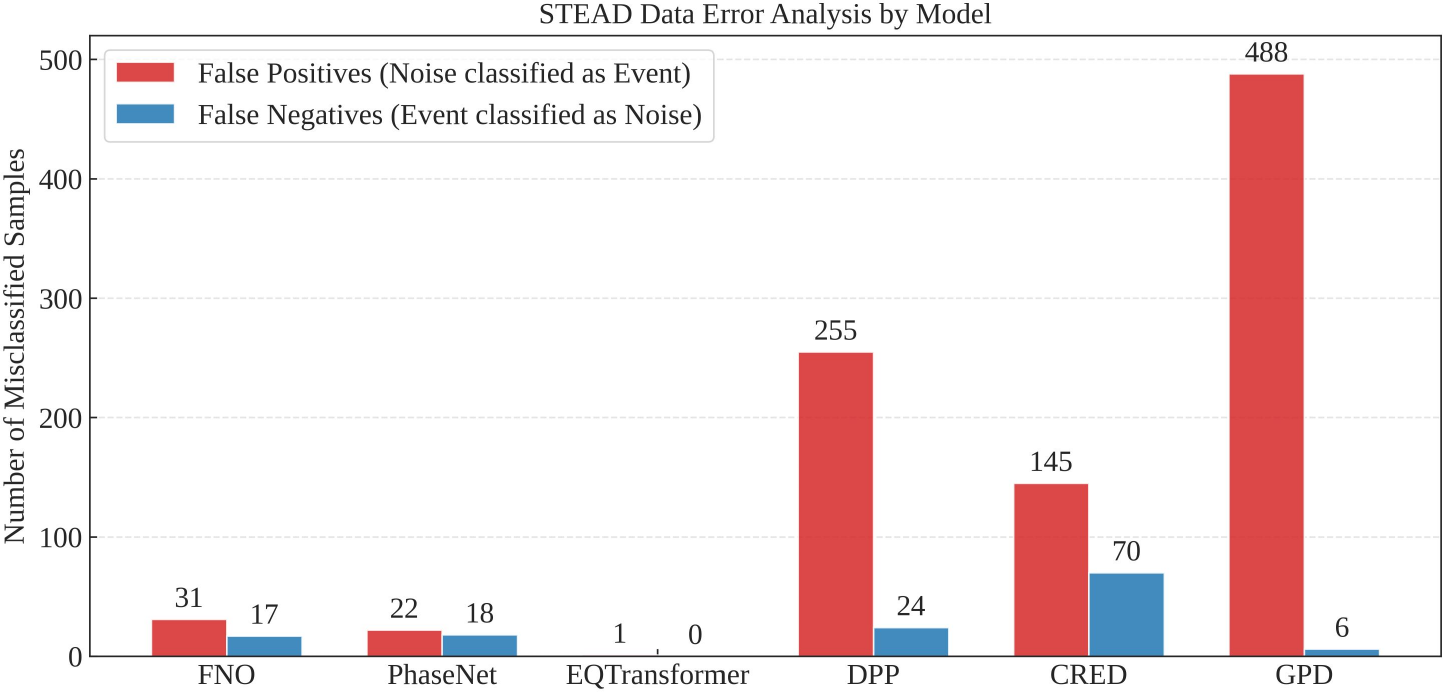}
\caption{
Error distribution analysis of the six DL models on the STEAD test set. The bar chart shows the False Positives (Type I errors, red) and False Negatives (Type II errors, blue). While PhaseNet and FNO demonstrate balanced and minimal error rates, CRED shows a high tendency for false alarms, whereas DPP is prone to missing valid seismic signals.
}
\label{fig:error_distr_comp}
\end{center}
\end{figure}

The error distribution in Figure~\ref{fig:error_distr_comp} further shows the performance of the different models. The FNO architecture maintains a balanced performance with 31 false positives and 17 false negatives. This suggests that the FNO architecture is not biased towards either class. In contrast, the CRED and DPP models show a high number of errors. These results confirm that the FNO model is a robust and efficient alternative for seismic event detection, achieving high human-like accuracy and generalization ability comparable to established benchmarks.

\begin{table*}[t!]
\centering
\caption{Classification performance metrics for the six deep learning models on the test set of 1,000 samples (500 noise, 500 signal) from the STEAD data. The best performing model in each category is highlighted in bold.}
\label{tab:model_performance}
\begin{tabular}{lcccccc}
\hline
\textbf{Model} & \textbf{Accuracy} & \textbf{Precision} & \textbf{Recall} & \textbf{F1 Score} & \textbf{False Positives} & \textbf{False Negatives} \\
\hline
EQTransformer & \textbf{0.999} & \textbf{0.998} & \textbf{1.000} & \textbf{0.999} & \textbf{1}   & \textbf{0}   \\
PhaseNet      & 0.960          & 0.956          & 0.964          & 0.960          & 22           & 18           \\
FNO           & 0.952          & 0.940          & 0.966          & 0.953          & 31           & 17           \\
CRED          & 0.785          & 0.748          & 0.860          & 0.800          & 145          & 70           \\
DPP           & 0.721          & 0.651          & 0.952          & 0.773          & 255          & 24           \\
GPD           & 0.506          & 0.503          & 0.988          & 0.667          & 488          & 6            \\
\hline
\end{tabular}
\end{table*}

\subsection*{Microseismic Data Performance}

\begin{figure}[!htbp]
\begin{center}
\includegraphics[width=0.5\linewidth]{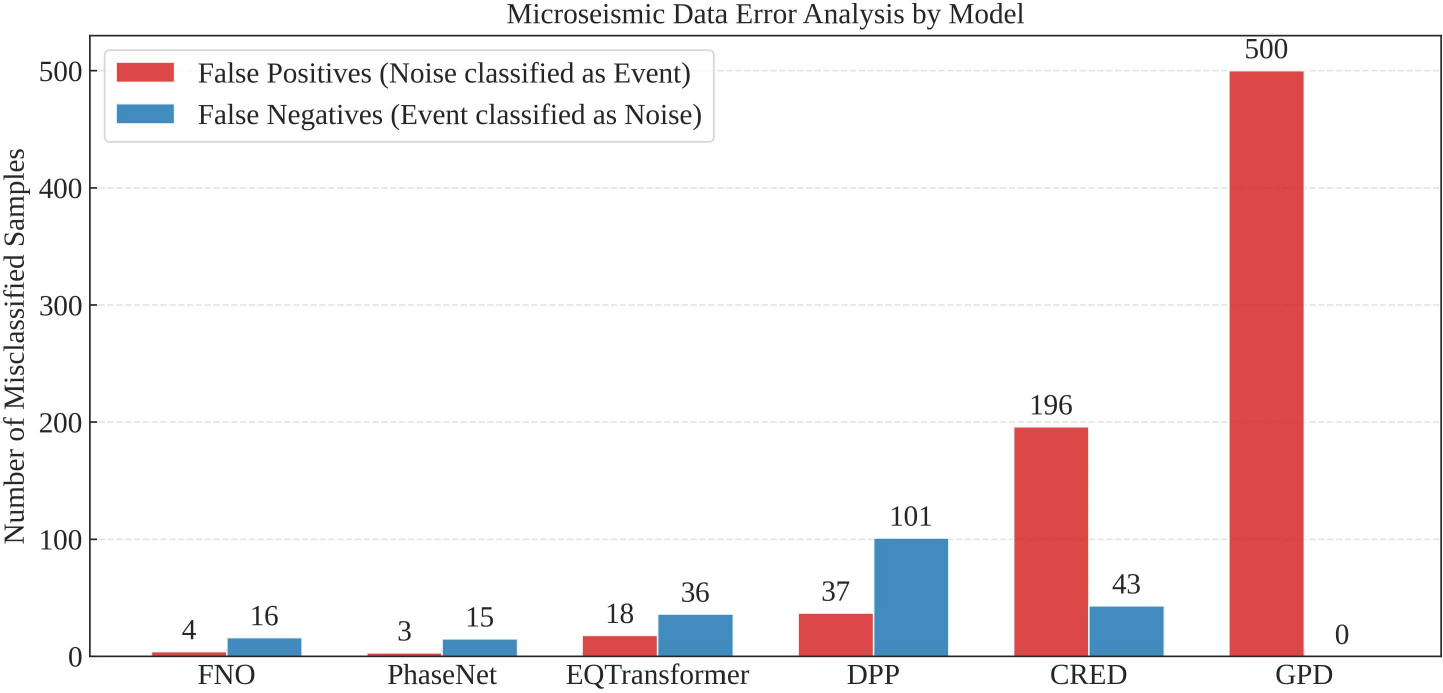}
\caption{
Error distribution analysis using microseismic test data for six models. The bar chart represents False Positives (Type I errors - red) and False Negatives (Type II errors - blue).
}
\label{fig:micro_error_distr_comp}
\end{center}
\end{figure}

In practical scenarios, models trained on one dataset often need to be applied to different, previously unseen datasets. Performance in such cross-domain applications may differ significantly from performance observed within the original training domain. These differences arise from variations in seismic data characteristics between the training and target sets and possibly some bias in the annotation process and the selection of the training data. To validate the performance of the model in different domains, a cross-evaluation was performed by testing each of the pretrained models on the microseismic test set (Table~\ref{tab:micro_model_perf}, Figure~\ref{fig:micro_error_distr_comp}).

\begin{table*}[ht]
\centering
\caption{Classification performance metrics for the six deep learning models on the blind test set of 1,000 samples (500 noise, 500 signal) from the microseismic data. The best performing model in each category is highlighted in bold.}
\label{tab:micro_model_perf}
\begin{tabular}{lcccccc}
\hline
\textbf{Model} & \textbf{Accuracy} & \textbf{Precision} & \textbf{Recall} & \textbf{F1 Score} & \textbf{False Positives} & \textbf{False Negatives} \\
\hline
FNO           & 0.980          & 0.992          & 0.968          & 0.980          & 4            & 16           \\
PhaseNet      & \textbf{0.982} & \textbf{0.994} & 0.970          & \textbf{0.982} & \textbf{3}   & 15           \\
EQTransformer & 0.946          & 0.963          & 0.928          & 0.945          & 18           & 36           \\
CRED          & 0.761          & 0.700          & 0.914          & 0.793          & 196          & 43           \\
DPP           & 0.862          & 0.915          & 0.798          & 0.853          & 37           & 101          \\
GPD           & 0.500          & 0.500          & \textbf{1.000} & 0.667          & 500          & \textbf{0}   \\
\hline
\end{tabular}
\end{table*}

Choosing the optimal probability threshold, our FNO-based model and PhaseNet achieved the highest scores (F1 = 0.98). EQTransformer was very close with an F1 of 0.95. Models with smaller input window size, such as DPP with an F1 of 0.85, CRED with an F1 of 0.79, and GPD with an F1 of 0.67, scored significantly lower. This highlights that our model, PhaseNet, and EQTransformer have an advantage due to their ability to analyze longer waveforms and have a broader context over time, making them perform better in event detection.

EQTransformer outperforms CRED because of the enhanced architecture of the model. EQT is an extended version of the CRED model that uses an additional attention component with CNN and RNN. DPP performs poorly because its hyperparameters were tuned for a local dataset that has very different seismic features. Although the other evaluated models were initially developed using local or regional datasets, their hyperparameters have not undergone the same level of systematic optimization as the DPP model, which may make them more robust when applied to diverse seismic data.

Differences in detection performance between the models may also stem from variations in their definitions of earthquake detections and the nature of the picks within the microseismic dataset. Specifically, GPD, PhaseNet, and DPP determine their detection scores directly from the noise probability, calculated as one minus the predicted noise likelihood. In contrast, CRED and EQTransformer produce explicit detection curves aligned with predefined detection labels that rely on both P- and S-phase annotations. Consequently, especially for regional datasets, these models declare detections only if both P and S annotations are available. In the microseismic dataset, however, certain waveform segments lack either P- or S-wave annotations. This incomplete labeling may particularly impact the detection accuracy of EQTransformer and CRED.

It is important to clarify that while this constitutes a cross-domain application, the source magnitudes in our target dataset ($0.5 < M < 2.5$) significantly overlap with the training distributions of most baseline models (typically $0.5 < M < 3.5$; see Table~\ref{tab:models}). Therefore, any performance degradation observed in the benchmark models is less likely attributable to a discrepancy in event magnitude and more likely driven by differences in sensor characteristics between the regional seismic networks and the local microseismic monitoring array.

\subsection*{Waveform Analysis and Error Characterization}

\begin{figure*}[!htbp]
\begin{center}
\includegraphics[width=0.65\textwidth]{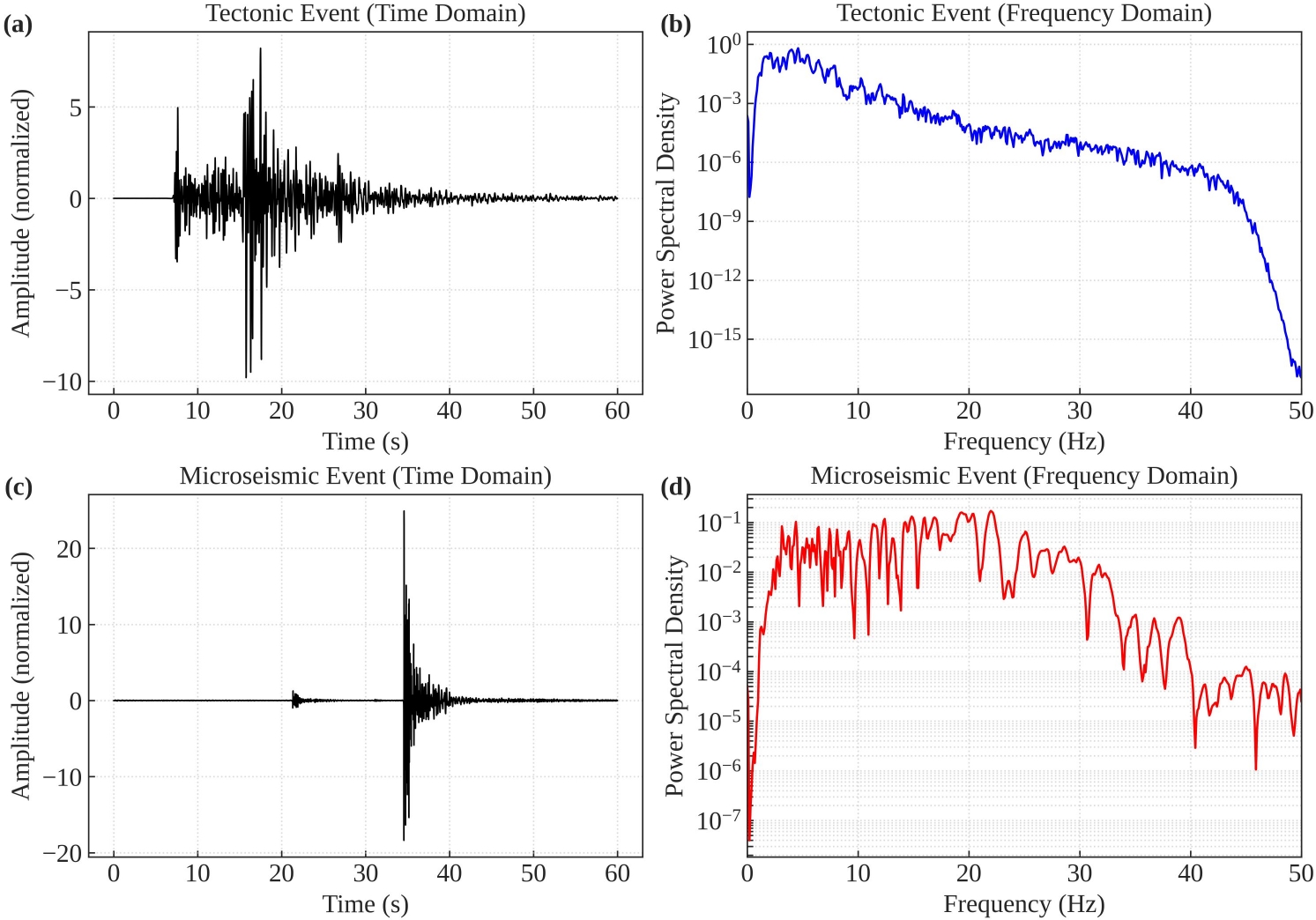}
\caption{
Comparison of example waveforms from the two datasets, illustrating their differing time–frequency characteristics. (a) STEAD tectonic earthquake. (b) Microseismic event from this study.
}
\label{fig:waveform_comp}
\end{center}
\end{figure*}

\begin{figure*}[!htbp]
\begin{center}
\includegraphics[width=0.65\textwidth]{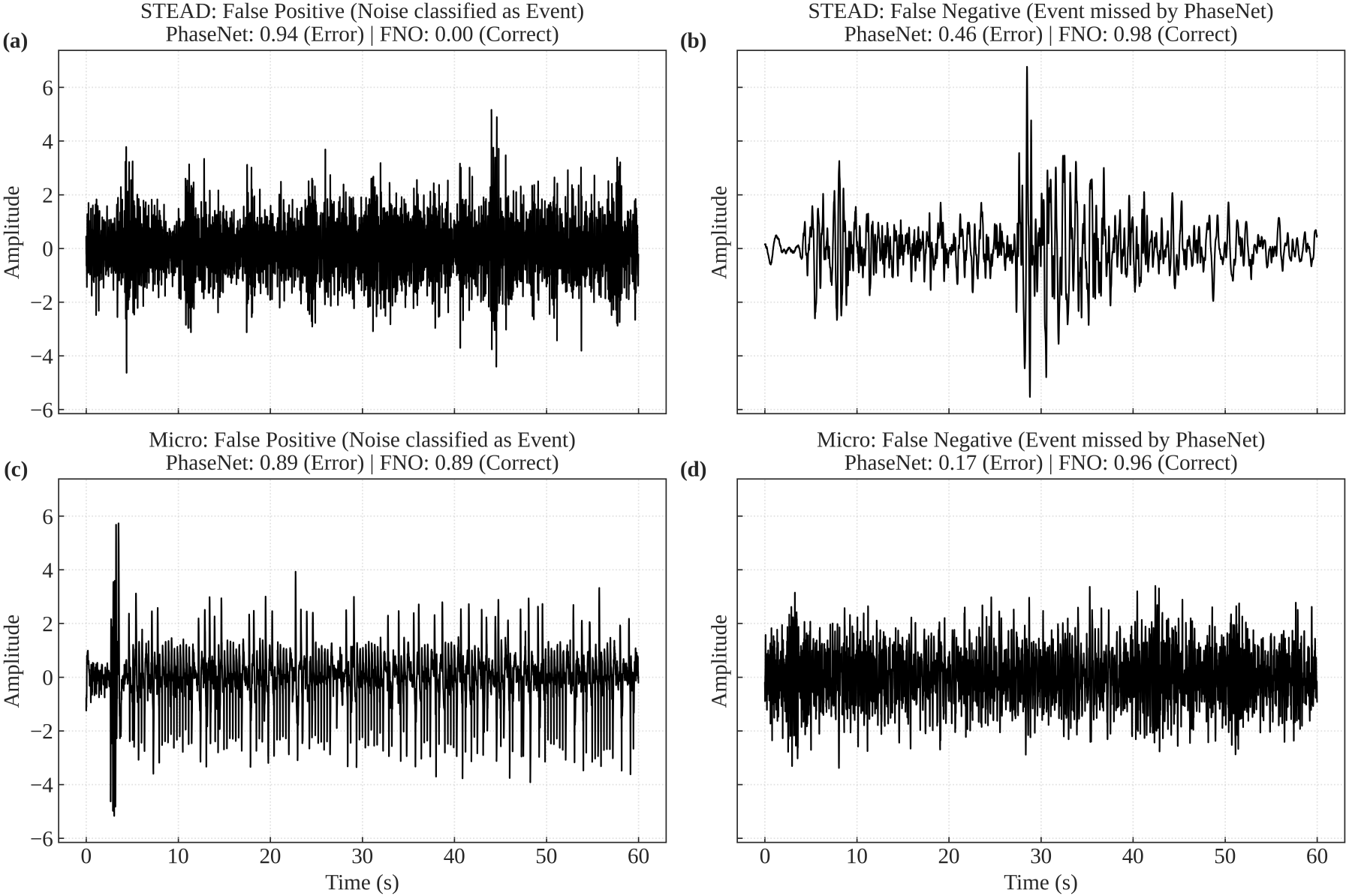}
\caption{
Analysis of classification failures by the baseline PhaseNet model on the STEAD and microseismic test sets. 
(a) A False Positive on STEAD, where instrumental noise is misclassified as a tectonic event. 
(b) A False Negative on STEAD, which indicates that an earthquake was missed. 
(c) A False Positive on the microseismic data. Although the models have similar probability scores for this noise data, PhaseNet identifies it as an event (above its optimal threshold $\tau = 0.85$). However, the FNO model rejects it, using a stricter threshold ($\tau = 0.90$).
(d) A False Negative on the microseismic dataset, where a low-magnitude event remains undetected. 
Prediction probabilities ($P$) for each model are indicated in the respective panel subtitles.
}
\label{fig:phasenet_fail}
\end{center}
\end{figure*}

To investigate the performance disparities observed in the statistical evaluation, we conducted a qualitative analysis of the waveform characteristics and specific failure modes. Figure~\ref{fig:waveform_comp} illustrates a domain mismatch between regional tectonic earthquakes in STEAD and locally recorded induced seismicity. The datasets differ systematically in recording scale (local vs. regional), propagation path/site response, and noise conditions, which together contribute to the observed performance gap. This discrepancy is a primary driver of the generalization failures observed in Figure~\ref{fig:phasenet_fail}, where the baseline PhaseNet model, optimized for longer-period tectonic events, may miss clear microseismic arrivals or misidentify high-frequency  noise resulting from anthropogenic activity as seismic signals.

\begin{figure*}[!htbp]
\begin{center}
\includegraphics[width=0.65\textwidth]{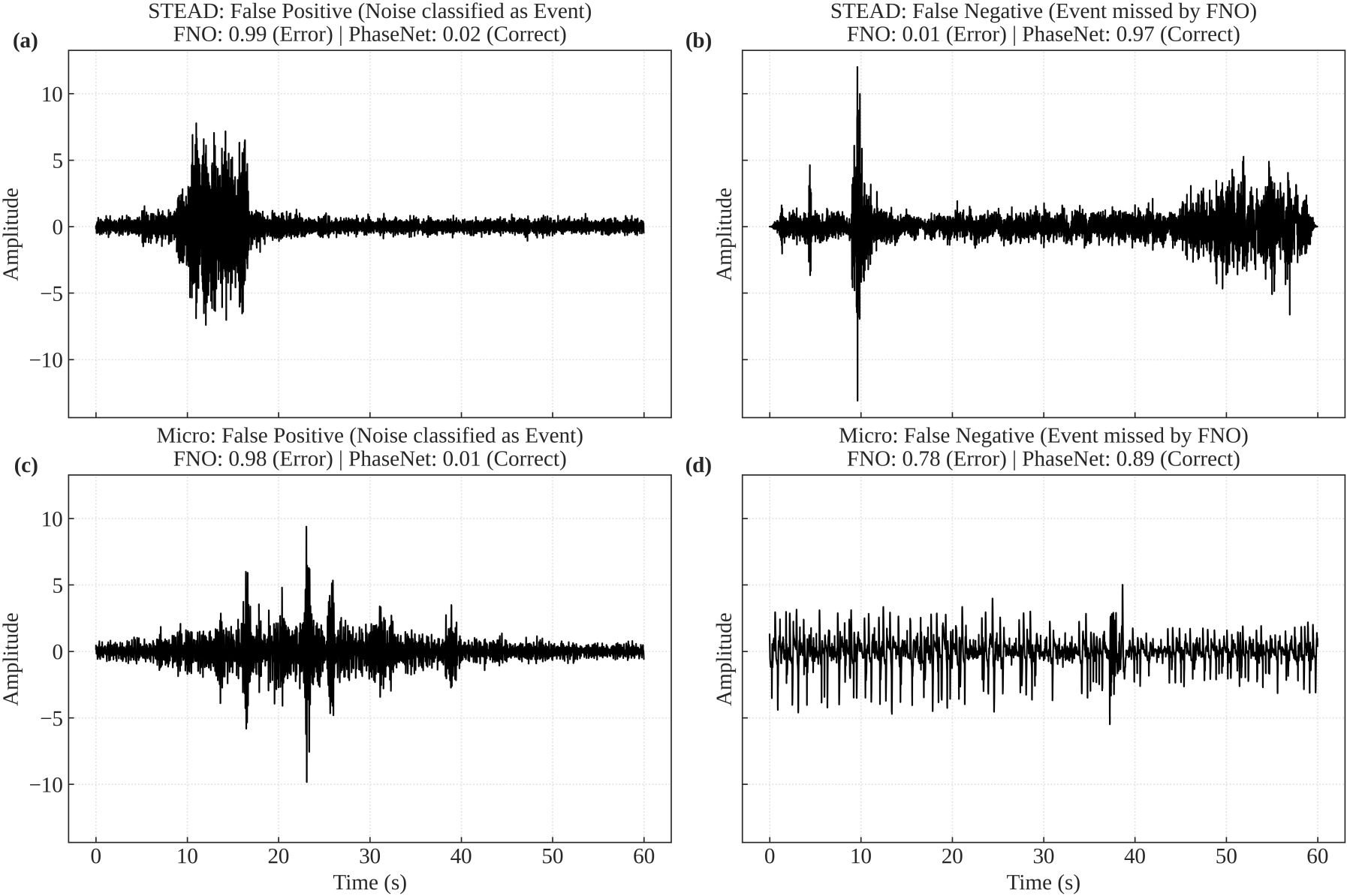}
\caption{
Analysis of classification failures by the proposed FNO model on the STEAD and microseismic test sets. 
(a) A False Positive on the STEAD dataset, where a complex noise burst is incorrectly flagged as a tectonic event. 
(b) A False Negative on STEAD, representing a missed earthquake signal, likely due to the model's limited capacity to resolve low-amplitude onsets within the coda of preceding events. 
(c) A False Positive on the microseismic dataset triggered by monochromatic resonance noise; the global spectral nature of the FNO architecture makes it sensitive to such strong periodic artifacts which mimic the spectral energy of a seismic signal. 
(d) A False Negative on the microseismic dataset, where a weak event ($SNR < 1.5$) is obscured by high-amplitude ambient field noise.
}
\label{fig:fno_fail}
\end{center}
\end{figure*}

In contrast, the failures of the FNO model (Figure~\ref{fig:fno_fail}) are less attributable to domain mismatch and more to the specific properties of the spectral processing. Because FNO layers rely on global Fourier transformations, they are highly efficient at capturing wave propagation patterns but can be susceptible to strong, monochromatic noise sources (e.g., resonance). In such cases (Figure~\ref{fig:fno_fail}c), the persistent spectral energy of the noise artifact overlaps with the learned frequency modes of seismic signals, leading to false positives. Furthermore, while FNO maintains high sensitivity, its lightweight architecture, comprising significantly fewer parameters than U-Net based models, may limit its ability to disentangle extremely weak signals from complex, non-Gaussian background noise (Figure~\ref{fig:fno_fail}d).

\subsection*{Multi-station Detection and SNR Effect}

\begin{figure*}[!htbp]
\begin{center}
\includegraphics[width=0.95\textwidth]{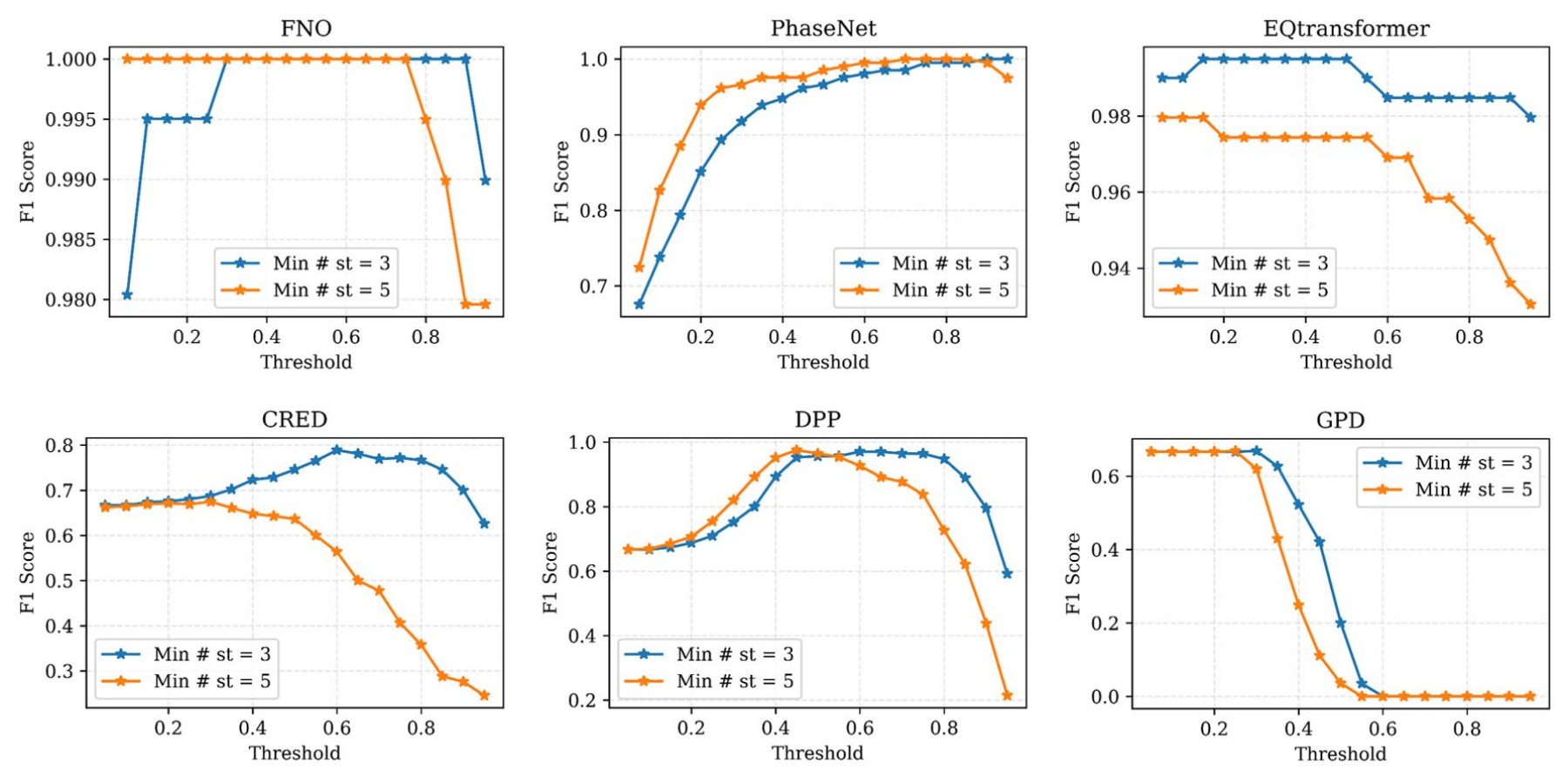}
\caption{
F1 score versus probability threshold results for the test set of 100 random triggered and 100 random false events from the microseismic dataset. Six deep learning models are evaluated. Trained on a limited number of data, the FNO-based model achieved an F1 score of 1.0.
}
\label{f1_vs_th_minst_all}
\end{center}
\end{figure*}

The deep-learning algorithms evaluated in this study are based primarily on single-station analysis strategies, which may lead to missed detections for events characterized by low signal-to-noise ratios (SNR) or false detections caused by local noise signals with emergent arrivals. To address this limitation and provide a more realistic assessment of practical performance, we additionally compared the models using a seismic array-based voting scheme (Figure~\ref{f1_vs_th_minst_all}). Specifically, detections were required to meet a minimum threshold of stations identifying the same seismic event. This approach simulates array-based event detection commonly employed by human analysts, who typically verify seismic events by observing consistent signals across multiple stations.

Generally, F1 scores increase with a higher threshold up to a value of threshold value where event detection falls off. The high threshold represents high certainty of detection, but it is not met by low signal events. This is consistent with the fact that the F1 declines faster for a minimum requirement of 5 stations than 3 stations. The increase in the F1 score for low thresholds results from a decrease in the number of false positives. We can conclude that the thresholds for network detection should be chosen based on the number of stations required for detection, as for each method, there is a 'sweet spot'. 

We observed significant differences in the F1 scores of different models evaluated on the microseismic datasets. This difference most likely stems from the inherent complexity of the detection task for each dataset and the methods used in dataset creation. For example, datasets with lower average signal-to-noise ratio (SNR) inherently present greater detection challenges than datasets with higher SNRs, which may degrade overall performance. Furthermore, inconsistencies in the dataset collection process, such as different quality control standards or annotations provided by multiple analysts, may lead to discrepancies that further decrease model performance. These observations highlight the need to evaluate all models on the same datasets in order to make direct and meaningful comparisons.

\begin{figure*}[!htbp]
\begin{center}
\includegraphics[width=0.95\textwidth]{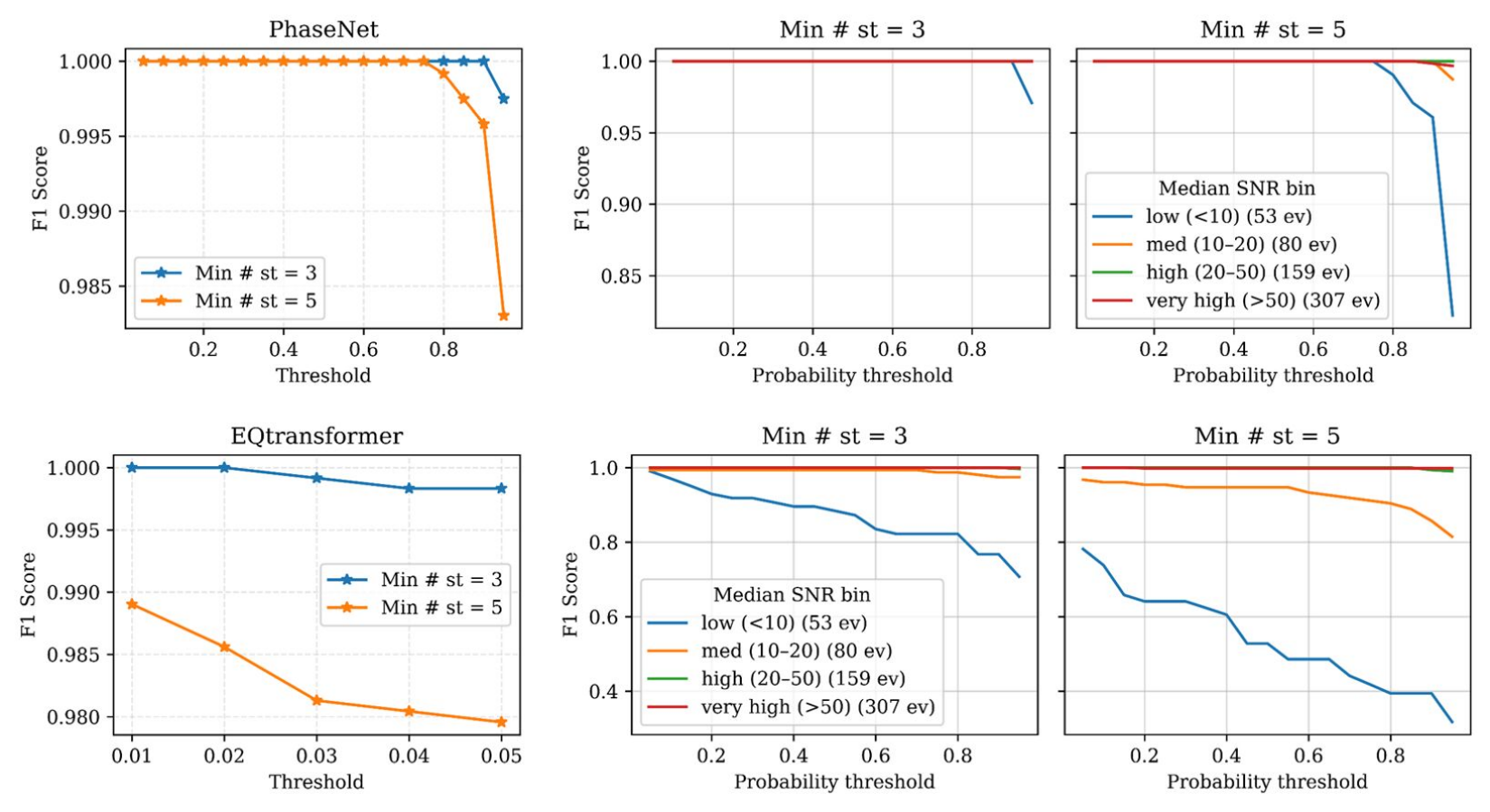}
\caption{
Influence of signal-to-noise ratio on the dependence of F1 score vs. probability threshold. The results are for all microseismic triggered events with 3 and 5 minimum stations voting for signal presence. Two deep learning models are evaluated.
}
\label{snr_vs_f1_vs_th}
\end{center}
\end{figure*}

Given the latter result, we compared the F1 scores of the different pickers with respect to SNR on our microseismic dataset (Figure~\ref{snr_vs_f1_vs_th}). We used the vertical component of traces for SNR calculation based on the P phase arrivals. As expected, we observe that pickers have an increasing F1 score with increasing SNR. These results suggest that EQtransformer is more sensitive to SNR than PhaseNet. Overall, the biggest gain of deep learning pickers compared to manual picking is their ability to pick successfully in low SNR conditions.

\subsection*{Computational Cost}

Deep neural networks such as EQTransformer and PhaseNet have significantly advanced earthquake detection and have been successfully deployed in real-time regional monitoring systems (e.g., the Southern California Seismic Network). However, optimization for strictly resource-constrained environments, such as downhole acquisition tools or battery-operated remote nodes, remains a valid engineering goal.

As shown in Table~\ref{tab:models}, PhaseNet and EQTransformer utilize approximately 268,000 and 377,000 parameters, respectively. In contrast, the proposed FNO-based model requires only ~34,000 parameters. This order-of-magnitude reduction in model size lowers the memory footprint and computational burden, making the FNO approach uniquely well-suited for edge-computing applications where hardware resources are severely limited, or for scenarios where models must be rapidly retrained on small, local datasets.

\section*{CONCLUSIONS}

We developed a lightweight Fourier Neural Operator (FNO) model for binary seismic-event classification from three-component waveform windows. The proposed architecture provides competitive detection performance while using an order of magnitude fewer trainable parameters than widely used deep-learning baselines (34,000 versus ~268,000 for PhaseNet and ~377,000 for EQTransformer), reducing memory footprint and inference cost. On STEAD, the FNO achieved F1 = 0.953 on 1,000 blind test windows and showed strong data efficiency, maintaining F1 > 0.93 even when trained on only 100 labeled examples. On a field microseismic dataset, the model reached F1 = 0.980 on 1,000 blind test windows after training on 200 labeled windows, closely matching PhaseNet (F1 = 0.982) and outperforming several benchmark models in cross-domain testing.

We further show that practical reliability can be improved through array-style decision-making: requiring detections to be supported by a minimum number of stations reduces false alarms, but necessitates careful selection of the probability threshold to balance missed events against false detections. Overall, the FNO-based classifier offers an efficient and accurate option for near–real-time microseismic monitoring, particularly in settings where computational resources and labeled training data are limited.

% \section*{ACKNOWLEDGEMENTS}

% We gratefully acknowledge the support of King Fahd University of Petroleum and Minerals (KFUPM) through the Outbound International Visiting Program grant no. ISP23224. 
% % We also extend our appreciation to Prof. Eric Verschuur and two anonymous reviewers for their insightful feedback, which significantly improved this manuscript.

\section*{AUTHOR CONTRIBUTIONS}

Conceptualization: Umair bin Waheed; 
Data curation: Ayrat Abdullin, Leo Eisner; 
Formal analysis: Umair bin Waheed, Leo Eisner, Abdullatif Al-Shuhail;
Methodology: Ayrat Abdullin, Umair bin Waheed, Leo Eisner; 
Project administration: Ayrat Abdullin; 
Resources: Umair bin Waheed, Leo Eisner; 
Supervision: Umair bin Waheed; 
Validation: Leo Eisner, Abdullatif Al-Shuhail;
Visualization: Ayrat Abdullin;
Writing - original draft: Ayrat Abdullin; 
Writing - review and editing: Ayrat Abdullin, Umair bin Waheed, Leo Eisner, Abdullatif Al-Shuhail.

\section*{DATA AVAILABILITY}

The data supporting the findings of this study are available from Seismik s.r.o., but restrictions apply to their availability. These data were used under license for the current study and are therefore not publicly available. Data are, however, available from the authors upon reasonable request and with permission of Seismik s.r.o.

\section*{COMPETING INTERESTS}

The authors declare no competing interests.

\section*{FUNDING}

The authors received no specific funding for this work.

\bibliography{ref_FNOdet_gp}

\end{document}